\documentclass[lettersize,journal]{IEEEtran}

\usepackage{amsmath,amssymb,amsfonts}
\usepackage{calc}
\usepackage{verbatim,leqno}
\usepackage[normalem]{ulem}
\usepackage{latexsym}
\usepackage{float,epsfig,color}
\usepackage{algorithm}
\usepackage{algorithmicx}
\usepackage{algcompatible}
\usepackage{algpseudocode}
\usepackage{subcaption}
\usepackage[bookmarks,breaklinks]{hyperref}
\usepackage{mdframed}
\usepackage{amsmath,amssymb,amsfonts}
\usepackage{calc}
\usepackage{verbatim,leqno}
\usepackage{latexsym}
\usepackage{float,epsfig,color}
\usepackage{blindtext}
\usepackage{graphicx}
\usepackage{bbm}
\usepackage{algorithm}
\usepackage{algcompatible}
\usepackage{multirow}
\usepackage{rotating}
\usepackage[nocomma]{optidef}
\usepackage{algpseudocode}

\newtheorem{lemma}{Lemma}

\ifCLASSOPTIONcompsoc
  \usepackage[nocompress]{cite}
\else
  \usepackage{cite}
\fi
\ifCLASSINFOpdf
\else
\fi
\begin{document}
%
\title{Linearized Data Center Workload and Cooling Management}
%
%
%
%

%

\author{Somayye Rostami\thanks{Somayye Rostami, Douglas G. Down and George Karakostas are with the Department of Computing and Software, McMaster Univercity, Hamilton, Ontario, Canada, (e-mails: rostas1@mcmaster.ca, downd@mcmaster.ca, karakos@mcmaster.ca)}, Douglas G. Down, \textit{Senior Member, IEEE}, George Karakostas}

\IEEEtitleabstractindextext{%
\begin{abstract}
With the current high levels of energy consumption of data centers, reducing power consumption by even a small percentage is beneficial. We propose a framework for thermal-aware workload distribution in a data center to reduce cooling power consumption. 
The framework includes linearization of the general optimization problem and proposing a heuristic to approximate the solution for the resulting Integer Linear Programming (ILP) problems.
We first define a general nonlinear power optimization problem including several cooling parameters, heat recirculation effects, and constraints on server temperatures. We propose to study a linearized version of the problem, which is easier to analyze. As an energy saving scenario and as a proof of concept for our approach, we also consider the possibility that the red-line temperature for idle servers is higher than that for busy servers. For the resulting ILP problem, we propose a heuristic for intelligent rounding of the fractional solution. Through numerical simulations, we compare our heuristics with two baseline algorithms. We also evaluate the performance of the solution of the linearized system on the original system. The results show that the proposed approach can reduce the cooling power consumption by more than 30 percent compared to the case of continuous utilizations and a single red-line temperature.

\end{abstract}
\def\abstractname{Note to Practitioners}
\begin{abstract}
   The approach proposed in this paper can be used as a baseline approach for tackling different power optimization problems for data centers. It proposes a general and efficient approach to solve the nonlinear problems especially when the closed-form functions, for example for the thermal model, are not available. A general nonlinear power optimization problem is defined. Linearization is then used to analyze the general problems and to propose a corresponding heuristic for approximation of the solution for ILP problems. The heuristic can be modified to solve similar problems. Scalability is an important factor that has been addressed in this approach because the linearized problem can be solved efficiently and calls to implicit functions are eliminated. So, this approach can be efficiently used for time sensitive calculations. The simulation results also show that the accuracy is maintained for larger problem sizes. For use in practice, better linearization approaches than regressing on a single linear function may be required. Our approach is adaptable to such settings. 
\end{abstract}

}

\maketitle

\IEEEdisplaynontitleabstractindextext

%
\IEEEpeerreviewmaketitle

\section{Introduction}
\label{int}






Data centers have been deployed to perform a large volume of computing tasks at
considerable operational costs. With the increasing demand for cloud computing and internet
services, data center operational costs have increased dramatically \cite{s1}. To reduce the power consumption of a data center, a number of techniques have been proposed at different levels including chip, server, rack and room levels \cite{s2}. This work focuses on thermal-aware workload distribution to reduce cooling power consumption, a main contributor to overall power consumption in a data center \cite{s1}.
Decreasing the power consumption by even a small percentage may generate significant energy savings. There have been a wide variety of optimization problems posed for such problems. Often, strong theoretical support is lacking. A general framework for systematic analysis in this area  would be desirable.

In terms of solution techniques, a deep learning approach is used in \cite{D1} and \cite{D2}. Proposing a heuristic (often a greedy heuristic) is also common \cite{tac}\cite{M1}\cite{jap}.    However, analysis of such heuristics is somewhat lacking. Some theoretical results are given in \cite{KKT}. Overall, we feel that a deeper analysis of the optimization approaches for such data center problems would be useful to guarantee an acceptable power reduction in different system instances. One way to do this is to adopt an approach that is standard in the area of control systems, for example. This involves linearizing a nonlinear system, with the idea that the linear system is easier to reason about. We find that this is the case in our setting, and we expect that this viewpoint can be of more general applicability. We believe that developing approaches to solve the resulting linear problem gives the promise of a single approach to solve a range of problems, rather than having to develop different heuristics for different (nonlinear) problems. If the decision variables are continuous, the linear problem can be solved by standard algorithms, otherwise if there are integral variables due to some energy saving or practical considerations, a single heuristic may address a range of similar problems. The linearization approach is especially efficient when closed-form functions are not available, for example evaluating the thermal model requires the numerical solution of a set of differential equations, which can become computationally quite expensive. Linear regression can then be used to estimate the functions. We also show that the proposed heuristic is more time efficient compared to popular approaches such as genetic algorithms (which typically also require ad hoc customization to perform well).

There are similarities and differences between the power optimization problems defined in the literature for thermal-aware workload distribution. The choices of power consumption model, performance and temperature constraints, and thermal models are the key points in the power optimization problems defined. Most of the problems are aimed at minimizing the total power consumption, consisting of the summation of IT and cooling power consumption. Different models for cooling power consumption are used, depending on which cooling facilities are present, i.e., CRAC units, fans, chillers \cite{tac}\cite{fan}\cite{sfan}. IT power consumption is also a function of utilization of a computing node. In some works, frequency scaling is considered so that IT power consumption is also a function of the operating frequency of a server \cite{sfan}\cite{pak}. Performance and temperature constraints are also commonly considered in the existing literature. In terms of performance constraints, it is typical to require that the summation of workload assignments to the servers (computing nodes) reaches a target demand level. There may also be a constraint on the maximum workload assigned to each server \cite{M1}\cite{KKT}\cite{pak}. 
For the temperature constraints, there is typically a red-line temperature that the inlet temperature of each server must not exceed.
Most of the literature assumes that the red-line temperature is a constant. However, in \cite{M1} the red-line temperature for each server is a function of its utilization. In terms of thermal models,  both steady-state and transient models are considered in the literature. Transient models are useful for real-time control settings \cite{MPC1}\cite{MPC2}\cite{MPC3}. Steady-state models are also common \cite{tac}\cite{M1}\cite{jap}\cite{sfan}.  With respect to thermal models, most of the literature considers the effect of heat recirculation on the server temperatures \cite{H1}. In this case, the inlet temperatures are a function of the servers' power consumption and the cooling parameters \cite{tac}. However, the models presented in \cite{M1} and \cite{sfan} do not consider recirculation effects which makes solving the optimization problem easier by removing the thermal interactions among the servers. In this work, a steady-state model with heat recirculation effects, several cooling parameters, and two red-line temperatures corresponding to idle and fully-utilized servers is considered.

 Server consolidation is another possibilty that is considered in the literature \cite{tac}\cite{jap}\cite{sfan}. The goal is to reduce the total idle power consumption by turning idle servers off. However, when turning idle servers off, the thermal model may change because the heat recirculation pattern among the servers changes. In this case, linear dependence of the temperatures on the servers' power consumption may be problematic. However, linear dependence is assumed in several papers, including \cite{tac} and \cite{jap}. In this work, we assume that idle servers are not turned off, which corresponds with practice. However, our approach can also be applied to the case of server consolidation.
 
 Heterogeneous and homogeneous data centers are considered in the literature \cite{tac}\cite{KKT}\cite{He1}\cite{He2}. Homogeneous data centers are a special case of heterogeneous data centers, where the servers are identical. In this case, the power optimization problem is simplified as can be seen in \cite{tac}.  If the number of servers chosen to be working is fixed, the total IT power consumption due to servers is a constant, leaving only the cooling power consumption to be minimized. In \cite{M1}, the data center model is heterogeneous and differences among the servers are highlighted. However, in \cite{KKT}, a homogeneous data center is considered as a practical case. In this work we focus on homogeneous data centers as a first step towards tackling more complicated problems.

 In this paper, we define a power optimization problem for a data center that is general enough to consider nonlinear dependencies and take into account different cooling parameters and heat recirculation effects. For our initial explorations, we make some assumptions. We assume that servers are identical and they are either idle or fully utilized. According to the temperature model in \cite{M1}, we study the effect of considering different server red-line temperatures as a function of server utilizations to see if exploiting this difference can reduce the total power consumption. To simplify these dependencies, we consider the case of having two red-line temperatures corresponding to idle and fully-utilized servers. In this case the power optimization problem becomes an integer programming problem. A physical model of a data center involves general nonlinear functions for the cooling power consumption and server temperatures. We then linearize these functions. After showing that the resulting linear optimization problem is NP-complete, we introduce simple rounding as a baseline approximation algorithm. Next, we introduce two problematic instances for simple rounding. An approximation scheme is then proposed to perform a more intelligent rounding. For comparison, a generic genetic algorithm is also introduced.  In the last section, we evaluate the proposed schemes for synthetic  systems (to test the limits of our heuristics) and a model developed from a working, experimental data center housed at McMaster University. We also evaluate the quality of the solution of the linearized problem by substituting it into the original, nonlinear model for the data center. Finally, as a concrete demonstration of our approach, we compare the case of considering two different red-line temperatures with the case of continuous utilizations but with one red-line temperature. Our contributions can be summarized as follows:
 
 \begin{itemize}
     \item Demonstrating the effectiveness of solving the general problem through linearizing the underlying thermal and power consumption models and formulating a corresponding optimization problem for the linearized system

     \item Theoretical analysis of the linear version including demonstrating that common approaches have the potential to be problematic
     \item Evaluation of the proposed heuristic on synthetic and practical instances 
     
     \item Evaluating the quality of the proposed solution approach by comparing with solving the original nonlinear problem directly
     
     \item Using the proposed approach to evaluate the energy savings possible for different red-line temperatures for idle and busy servers
     
 \end{itemize}
 
The remainder of the paper is organized as follows. Section \ref{sys} gives the underlying nonlinear system model and corresponding power optimization problem. Section \ref{def} discusses the linearized version of the nonlinear problem. Section \ref{app} is concerned with the solution of the linearized problem. Section \ref{eva} provides experimental evaluation of our approach, both in terms of the limits of our heuristics and  the applicability to a problem using a model from an operational data center. Finally, Section \ref{con} provides concluding remarks and future research directions.

\section{System Model and Motivation}
\label{sys}

Our starting point is a general power optimization problem introduced in \cite{M3}. We consider $n$ servers located in racks and cooled with a cooling system parameterized by $m$ adjustable cooling parameters, i.e., CRAC units' reference temperatures, air flows, and chilled water temperature. The cooling parameters are denoted by $v_1, v_2, ..., v_m$.
The aim is to minimize the total power consumption for the data center consisting of the summation of cooling and IT power consumption. The cooling power consumption is a function $F$ of the values of cooling parameters. The IT power consumption is a function of server utilizations, denoted by $\rho_1, \rho_2,..., \rho_n$. So, the decision variables are $v_1,..., v_m$ and $\rho_1,..., \rho_n$. We will also represent the decision variables by the vectors $v$ and $\rho$. We assume that the servers are identical.

The cooling variables are continuous and have a lower and upper bound given by $V_{LB}^{(j)}$ and $V_{UB}^{(j)}$ for $v_j$, $j=1,...,m$, i.e., input vectors $V_{LB}$ and $V_{UB}$ are of size $m$. 
For the utilization, we assume that servers are either idle or fully-utilized, so that the utilization of server $i$, denoted by $\rho_i$, is 0 or 1, $i=1,...,n$. However, to satisfy performance constraints, the maximum utilization of each server may be required to be strictly less than 1. This case can be easily transformed to a similar 0-1 optimization problem.

While minimizing power consumption, thermal constraints must also be respected to ensure reliable operation of servers. The inlet temperature of a server should not exceed a red-line temperature. The vector of inlet temperatures is denoted by $T$ with size $n$. We consider a steady-state model for the inlet temperatures. In this case, the inlet temperatures depend on the value of cooling parameters and the servers' IT power consumptions (which are a function of server utilizations). This dependence is given by the function $M(v, \rho)$. Referring to \cite{M1}, the red-line temperature for a server can also be a function of its utilization. In fact, more lightly utilized servers can have higher red-line temperatures. As we have assumed that server utilizations are either 0 or 1, we consider two red-line temperatures for each server, $T_{idle}$ and $T_{busy}$, corresponding to the utilizations 0 and 1, respectively. According to the model presented in \cite{M1}, $T_{idle} > T_{busy}$. This allows a lower cooling effort for idle servers, as they can tolerate a higher temperature.

There is also a performance constraint, related to quality of service. Because the servers are either idle or fully utilized and have the same computational capacities, the performance criterion is translated to a required number of servers working, denoted by $D$. This means at least $D$ servers should be working (note that the value of $D$ may be higher than required to support desired throughput, for example to satisfy latency requirements). As an increase in the number of busy servers tends to increase the inlet temperatures resulting in higher cooling and IT power consumption, the performance constraint is tight for the optimal power consumption, so exactly $D$ servers are working. In this case, the IT power consumption is fixed and minimizing the cooling power consumption is equivalent to minimizing the total power consumption.

Based on the preceding discussion we can formulate the following optimization problem:

\begin{mini!}|s|[2]                
{}
{F(v) \label{P0} \tag{P0}}
{\label{k}}
{}
\addConstraint{\sum_{i=1}^{n} \rho_i }{\geq  D \label{c3}}
\addConstraint{ M(v,\rho)}{\leq  T_{idle} 1_{n \times 1} - (T_{idle}-T_{busy})\rho  \label{c7}}
\addConstraint{ v}{ \geq V_{LB}\label{c4}}
\addConstraint{v}{\leq  V_{UB}\label{c5}}
\addConstraint{ \rho_i}{ \in  \{0,1\} \quad \forall i=1,...,n\label{c2}}
\end{mini!}
 where $F(v)$ is the cooling power consumption corresponding to the cooling variable vector $v$ and $M(v,\rho)$ is the function corresponding to the thermal model. Constraint \eqref{c3} is the performance constraint, and constraint \eqref{c7} limits the inlet temperatures to be less than the corresponding red-line temperatures. 
As explained, the performance constraint is tight for an optimal solution and, therefore, we need to decide which $D$ servers should work to minimize the cooling power consumption while satisfying the temperature constraints.

As an example, we consider the data center modeled in \cite{Roh}, an experimental data center housed at McMaster University. Similar to their model, there are five racks. Each rack consists of five servers. So, there are 25 servers in total. The cooling variables are the chilled water temperature and total air flow generated by two fans located at both ends of the five racks (the air flow is divided equally between the two fans). Thus, there are two cooling variables.\ Fig.\ 1 shows the top view for this system. To calculate the servers' temperatures in terms of the two cooling variables and the server utilizations, we use a model from \cite{Roh}. This is the function $M$ in problem \eqref{P0}. $M$ is not explicitly given and its calculation is based on numerically solving a set of differential equations, so it is expensive to be called. For the function $F$ in problem \eqref{P0}, we also use the models developed in \cite{Roh} for fan and chiller power consumption. We first tried to solve problem \eqref{P0} for this system using MATLAB (the \textit{surrogateopt} function). The platform was MATLAB R2021b running on a 64-bit system with an i7-1185G7 processor and 8-GB RAM. Using the default settings of the MATLAB function, the number of iterations to search the feasible space of the problem is more than 1000 and each iteration takes about 1.4 seconds to be run. The bulk of the execution time is the calculation of the function $M$ at each iteration. So, using MATLAB is not efficient to solve the problem. In particular, as we see in Section 5, the solution returned by MATLAB after a limited number of iterations may not be close to optimal. This inefficiency is the main motivation for our proposed approach, which is explained in the following sections.

\begin{figure}\label{opt}
\includegraphics[width=9cm]{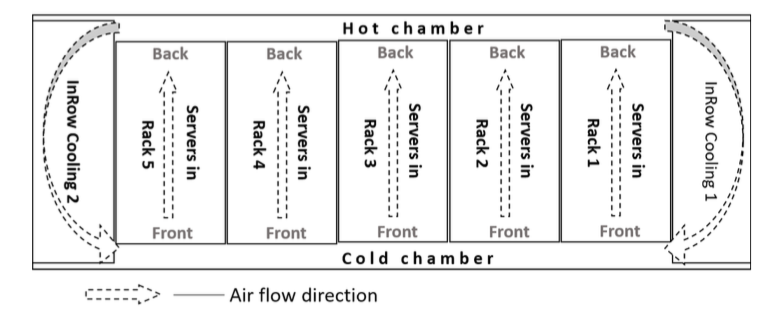}
\centering
\caption{The data center's top view according to \cite{Roh}\cite{M2}}
\end{figure}

\section{Problem Definition}
\label{def}

To find efficient heuristics to approximate the solution of problem \eqref{P0}, we first make some simplifying assumptions, which allow us to transform problem \eqref{P0} to the version of the problem considered in this paper. The assumptions are used for developing heuristics in Section \ref{heu}.

\begin{itemize}

\item Our main assumption is that $F$ and $M$ are linear functions. This allows us to achieve two goals. The first is to check the validity of the following procedure: performing regression and linearizing the system, solving the linear optimization problem and using the solution for the original system. A secondary goal is to assess the difficulty of problem \eqref{P0} by first studying its simplest version.
	\item $F$ is an increasing function of the cooling variables. This is without loss of generality for two reasons. First, the cooling power consumption is monotone with respect to cooling system parameters such as fan speeds and reference temperatures of CRAC units. Second, for some variables such as reference temperatures of CRAC units, the function $F$ is decreasing, however by transforming the corresponding cooling variables, we can convert $F$ to an increasing function.
	 
		 \item $M$ is a non-increasing function in terms of the cooling variables. This follows from the fact that to decrease the inlet temperatures, the cooling power consumption must be increased by increasing the value of cooling variables (recall that $F$ is an increasing function).
		 \item $M$ is a non-decreasing function in terms of the server utilizations. This means when a server's utilization increases (equivalent to an increase in the server power consumption), the inlet temperatures also increase. This is also a reasonable assumption in practice.

	\item The problem is feasible for every vector $\rho$, i.e., if we fix the vector $\rho$, the feasible space for $v$ is not empty. This is a result of the assumption that, in practice, maximizing the cooling effort is enough to prevent all of the servers from exceeding their red-line temperatures.

\end{itemize}

Using our assumptions, the following notation will simplify our exposition: 
\begin{itemize}
\item $M= -A v+ B\rho + E$, where $A_{i,j}, B_{i,j} \geq 0$ (all entries are non-negative). $A_{n \times m}$ is the cooling matrix, $B_{n \times n}$ is the heat recirculation matrix and $E_{n \times 1}$ is the constant part. $B_{i,j} \geq 0$ allows the optimal solution to have exactly $D$ servers working. $A_{i,j} \geq 0$ will allow us to determine the dominant cooling variables in the proposed heuristic in Section \ref{heu}.
\item  $b=T_{idle}$, and
$a=T_{idle}-T_{busy}>0$ 
\item The cooling power consumption is $F=\sum_{j=1}^{m} c_jv_j$, where the coefficients $c_j$ relate the value of cooling parameters to the power consumption of the corresponding cooling facilities. 
We normalize the cooling variables so that $F= \sum_{j=1}^{m} v_j$. This changes the values of $A$, $V_{LB}$ and $V_{UB}$ but for convenience we use the same notation for the formulation.
\end{itemize}

As a result, our problem can be formulated in scalar form as follows: 

\begin{equation}\label{p1}
		\begin{aligned}
		 {\min} \quad  \sum_{j=1}^{m}  v_j &   \\
			  \text{s.t.} \quad			 
			\sum_{i=1}^{n} \rho_i & \geq D \\
						-\sum_{j=1}^{m}A_{l,j} v_{j}{+} 
      	\sum_{i=1}^{n} B_{l,i} \rho_i 	+ a \rho_l 		& \leq  b- E_l   \quad \forall l=1,...,n  \\			 
			v_{j} & \geq  V_{LB}^{(j)}  \: \: \: \quad  \forall j=1,...,m  \\
			v_{j}  & \leq  V_{UB}^{(j)}  \: \: \: \quad  \forall j=1,...,m  \\
												 \rho_i & \in  \{0,1\} \:  \quad   \forall i=1,...,n \\
		\end{aligned}
	\end{equation}
	
	As we will see in Section \ref{eva}, solving this problem does not scale well using existing packages (in our case MATLAB). This is the motivation for proposing approximation schemes in the next section.

\section{Approximation of the Optimal Solution}
\label{app}
We first show that problem \eqref{p1} is NP-complete. We then introduce simple rounding as a base algorithm to approximate \eqref{p1}. We show that simple rounding may be problematic by constructing two bad cases for the algorithm. Then we propose an approximation scheme to perform more intelligent rounding. Finally, we present a generic genetic algorithm as a standard approximation approach. The main purpose of presenting simple rounding and a genetic algorithm is to better understand how well the proposed scheme performs as compared to standard schemes.

\subsection{NP-completeness}

To show the NP-completeness of the problem, we assume a particular instance where $m=1$, $A=1_{n \times 1}$ and $E=0_{n \times 1}$. In this case, there is only one cooling variable, denoted by $v$. We also set $V_{LB}^{(1)}= 0$ and $V_{UB}^{(1)}= \infty$. Then the problem becomes





 
 \begin{equation}\label{neww}
		\begin{aligned}
		 \min \quad  v \quad \quad & \\
			\text{s.t.} \quad
			\sum_{i=1}^{n} \rho_i &\geq  D\\						 -v+ \sum_{i=1}^{n}B_{l,i} \rho_i +a\rho_l &\leq  b  \quad \quad \quad \forall l=1,...,n\\	\rho_i & \in   \{0,1\} \quad \forall i=1,...,n \\
								v & \geq  0  \\
						\end{aligned}
	\end{equation}
	
	An equivalent formulation of \eqref{neww} is
	



	
	\begin{equation}\label{new22}
		\begin{aligned}
		 \min  \quad  \max \: \: & B'\rho \\
			\text{s.t.} \quad
		\sum_{i=1}^{n} \rho_i & =  D \\
				\rho_i & \in   \{0,1\} \quad \forall i=1,...,n \\			
		\end{aligned}
	\end{equation}
			where $B'=B+ a I_{n \times n}$ and $I$ is the identity matrix. In \cite{tac}, problem \eqref{new22} is shown to be NP-complete by a reduction to a partitioning problem. 
			
	\subsection{Bad Cases for Simple Rounding}
	
	We first consider simple rounding as a baseline approach. In this scheme, we solve the LP relaxation of \eqref{p1} (by allowing the utilizations to be fractional, $0 \leq \rho_i \leq 1$) and round the $D$ largest values to 1 and the remaining values to 0. We will see that simple rounding may have unacceptable performance. To show this, let us consider two special instances of \eqref{p1}:
	
	\begin{itemize}
		\item \textbf{Case 1:} There are $p \leq n$ entries equal to 1 for each row of $B$; the remaining entries are 0. For each row of $A$ one entry is equal to $q>0$ and the others are 0. Also, $V_{LB}^{(j)}= v_{L}$, 
 $\forall j=1,...,m$, and $E=0_{n \times 1}$.
		\item \textbf{Case 2:} In $B$ the entries are 0 or 1. If $A_j \neq A_i$, where $A_i$ is the $i$th row of $A$, then $B_{i,j}= B_{j,i}=0$ which means that servers $i$ and $j$ are \textit{isolated} from each other. Assume $S$ is a set of non-isolated servers (those with equal corresponding rows in $A$), then for $B_S$ (the sub-matrix of $B$ corresponding to the servers in $S$) the number of ones in each row and each column is $p$. Finally, $E=0_{n \times 1}, V_{LB}=0, V_{UB}= \infty$.
			\end{itemize}

\begin{lemma}
 If in Case 1,  $\frac{D}{n}(p+a) \leq b +  q v_{L}$, then there is an optimal solution for the LP relaxation of \eqref{p1} that distributes the load $D$ uniformly among the servers and satisfies $v^*=V_{LB}$. In this case, simple rounding does not guarantee a bounded approximation factor (the ratio of the solution found by the algorithm to the optimal solution).
\end{lemma}

\begin{IEEEproof}
The inequality $\frac{D}{n}(p+a) \leq b +  q v_{L}$ shows that in Case 1 by distributing the load $D$ uniformly and setting $v=V_{LB}$ we can satisfy all the constraints in the problem. So an optimal solution is $v^*=V_{LB}, \rho^*= \frac{D}{n} 1_{n \times 1}$ and $c^*=mv_L$, where $c^*$ is the optimal fractional cost.
With respect to the approximation factor, it is enough to show that it is not bounded for one specific example of Case 1. We define $\hat{B}$ as follows. We set $\hat{B}_{i,i}=1, i=1,...,D$, while the remaining entries that are set to one (each row has a total of $p$ ones according to the definition of Case 1) are chosen from the columns $D+1$ to $n$. We show that when $B=\hat{B}$, the ratio of the approximate cost to the optimal integral cost can grow arbitrarily large.

Because of symmetry, for simple rounding it does not matter which $\rho^*_i$ values are rounded up to 1. Assume that the $\rho^*_j$ values with $B_{i,j}=1$ for some row $i$ in $B$ are rounded. Depending on which of $D$ or $p$ is smaller, $D$ or $p$  $\rho^*_j$ values are rounded, respectively (potentially including $\rho^*_i$). Now suppose that

\begin{equation} \label{p2}
	\frac{D}{n}(p+a) \leq b +  q v_{L} < min(D, p)+a, 	\end{equation}
and define $ s= \frac{min(D, p)+a}{b +  q v_{L}}$ ($s>1$ because of the second inequality). To satisfy the constraints in \eqref{p1} for the rounded $\rho^*$, we must have $b+ qv' \geq (b +  q v_{L})s= b+(sq+ \frac{(s-1)b}{v_{L}})v_{L}$, where $v'$ is the value of the corresponding cooling variable needed for cooling server $i$ (that with non-zero entry in the $i$th row of $A$). So, $\frac{v'}{v_{L}} \geq \frac{(s-1)b}{qv_{L}}$ and

\begin{equation}
    \frac{\hat{c}}{c^*} \geq \frac{v'+(m-1)v_L}{mv_L} \geq \frac{v'}{mv_L} \geq \frac{(s-1)b}{mqv_{L}}
\end{equation}
where $\hat{c}$ is the cost of the solution for simple rounding. By reducing $v_{L}$ or $q$ in $\frac{(s-1)b}{mqv_{L}}$ (while maintaining \eqref{p2}, for example by increasing $n$), $\frac{\hat{c}}{c^*}$ can grow unboundedly. For example, set $p=3, a=1, b=2, q=1, D>p, m=1$, then $\frac{\hat{c}}{c^*}$ is at least $\frac{2(2-v_L)}{(2+v_L)v_L}$, which can be made arbitrarily large by decreasing $v_L$ (and decreasing $\frac{D}{n}$ to maintain \eqref{p2}).  We show that the optimal integral solution can also be equal to $c^*$.
It is possible to have a total load of at least $D$ by choosing at most $\lfloor b+qv_{L} - a \rfloor$ servers in each row of $B$ (the specific choice depends on the structure of $B$), hence the optimal fractional and integral solutions are equal in terms of cooling variable values (setting the cooling variables equal to $V_{LB}$ is enough to cool the servers). Let us set $b-a=1$ ($\lfloor b+qv_{L} - a \rfloor \geq 1$) and $B=\hat{B}$.  By choosing the servers $1,...,D$, at most one server is chosen at each row of $B$. 
\end{IEEEproof}

\begin{lemma}
 In Case 2, there is an optimal solution for the LP relaxation of \eqref{p1} that distributes the load uniformly among the servers that are not isolated. In this case, simple rounding does not guarantee a bounded approximation factor. 
\end{lemma}

\begin{IEEEproof}
In Case 2, assume that $S$ with size $n' \leq n$ is a set of non-isolated servers with the corresponding row $A_i$ in $A$. If the total load for $S$ is $D'$, it is not difficult to see that in this case distributing the load $D'$ uniformly among the servers in $S$ is an optimal workload distribution because based on the conditions for Case 2, the average of the values in $B_S \times \rho_S$ is fixed and equal to $\frac{D'p}{n'}$ (for the numerator, each element of $\rho_S$ is repeated $p$ times in the summation because there are $p$ ones in each column of $B_S$), and for optimality we want to minimize the maximum of the values in $B_S  \rho_S$. To be more precise, $max(B_S  \rho_S+a \rho_S)-b$ determines the increase needed for the cooling variables to cool the servers in $S$. Similar to Lemma 1, we show that the approximation factor can be made arbitrarily large for $B=\hat{B}$, where in the $i$th row of $\hat{B}$, the $i$th to $(i+p-1)$th (mod $n$) entries are 1 (this is according to the definition of Case 2). For the proof, we assume $n'=n$ and the rounding is similar to what was done in the previous lemma. When $n'=n$, the optimal cost for $\rho$ is proportional to $max(B  \rho+a \rho)-b$, i.e., $c^*(\rho)= \alpha (max(B  \rho+a \rho)-b)$ where $\alpha$ is a constant (if $max(B  \rho+a \rho) \leq b$ then the optimal cooling variables are $V_{LB}=0$). We show that the ratio of the approximate solution to the optimal integral solution can grow unboundedly. Note that according to the rounding in Lemma 1 and proportionality of the optimal cost to the max term we explained earlier, the optimal cost for the rounded solution, $\hat{c}$, is proportional to $min(D, p)+a-b$. Let us set $\frac{D}{n}=\frac{1}{p}, p<D$ and $B=\hat{B}$.  If we choose the $1$st, $(p+1)$th, $(2p+1)$th, ..., $((D-1)p+1)$th servers, then in each row exactly one server is chosen and the optimal integral cost, $c^*$, is proportional to $1+a-b$ (assuming that $1+a>b$). So $\frac{\hat{c}}{c^*}=\frac{p+a-b}{1+a-b}$, which can be made arbitrarily large by increasing $p$. 

\end{IEEEproof}

\subsection{Proposed Approximation Schemes}
\label{heu}

We propose two variants of an approximation scheme for problem \eqref{p1}. The approximation schemes are based on assumptions that are appropriate for practical systems. The other approach is a standard genetic algorithm which is used for comparison purposes.
\subsubsection{Maxmin Scheme}

In this section, we propose a more intelligent rounding than simple rounding. Let us denote the optimal fractional solution for the LP relaxation of \eqref{p1} by $(\rho^*, v^*)$. In the proposed scheme, instead of just rounding $\rho^*$, the values in $B' \rho$ (recall that $B'=B+I_{n \times n}$ where $I$ is the identity matrix) are also considered. Assume $\hat{\rho}$ is an integral workload distribution with total load $D$. If we replace $\rho^*$ with $\hat{\rho}$, and set the cooling variables equal to $v^*$, the temperature constraints for some servers are violated. Thus, some values in $v^*$ must increase to compensate for these violations. The sizes of the violations depend on the values in $B' \hat{\rho}$. We compute the violations for $\hat{\rho}$ as follows:

\begin{equation}
	(\sum_{i=1}^{n}B'_{l,i} \hat{\rho}_i -(\sum_{j=1}^{m}A_{l,j} v^*_{j}+b-E_l) )^+\quad l=1,...,n\end{equation}

For the power consumption optimization, the total increase in $v^*$ must be minimized. Now, if the maximum violation is small for some vector $\hat{\rho}$, the total increase needed in $v^*$ to compensate should also be small. For the other violations corresponding to $\hat{\rho}$, there are two possibilities: they are much smaller than the maximum violation or they are on the order of the maximum violation. The first case can be fixed by a small increase. In the second case, due to correlations, we would expect at least some of the violation to be corrected when the maximum violation is addressed. In this case fixing the maximum violation will partly fix the other correlated violations. Two servers $i$ and $j$ are correlated if there are similarities between their corresponding rows in $A$ and $B'$ (servers that are near to each other may be correlated). Thus, with a modification (considering weights $w_l$ defined below), the problem we solve is:

\begin{equation}\label{new1}
		\begin{aligned}
		 \min  \quad \max \limits_{ 1 \leq l \leq n } & \frac{[\sum_{i=1}^{n}B'_{l,i} \rho_i-(\sum_{j=1}^{m}A_{l,j} v^*_{j}+b-E_l)]}{ w_l}  \\
			\text{s.t.} \quad
			\sum_{i=1}^{n} \rho_i & =  D\\			\rho_i & \in   \{0,1\} \quad \forall i=1,...,n \\			
		\end{aligned}
	\end{equation}
where $w_l$ is the maximum entry in the $l$th row of $A$. The index of $w_l$ in the $l$th row of $A$ determines the dominant cooling variable corresponding to server $l$. The $w_l$ values translate the value of violations to the increases needed in the cooling variables. We need this translation to compare the violations in terms of the increase needed to compensate. To be more precise, we assume that a violation is mostly compensated for by the dominant cooling variable for each server and the ratio in the cost function measures the increase needed in this case. 


However, the assumption that the violations that are on the order of the maximum violation are correlated may not be true in general. Problem \eqref{new1} minimizes the maximum increase needed among all dominant cooling variables. This means it considers the maximum increase instead of the total increase in all dominant cooling variables. If there are $K \leq m$ dominant cooling variables, denoted by $d_k$, $k=1,...,K$, and $S_k$ is the set of servers with dominant cooling variable $d_k$, the problem corresponding to minimizing the total increase is the following:

\begin{equation}\label{new2}
		\begin{aligned}
		 \min \quad \sum_{k=1}^{K} & \max \limits_{ l \in S_k} \frac{[\sum_{i=1}^{n}B'_{l,i} \rho_i-(\sum_{j=1}^{m}A_{l,j} v^*_{j}+b-E_l)]^+}{ w_l}  \\
			\text{s.t.} \quad
			\sum_{i=1}^{n} & \rho_i  =  D\\		
			& \rho_i \in   \{0,1\} \quad \forall i=1,...,n \\			
		\end{aligned}
	\end{equation}

	The max term is for calculating the increase needed for the cooling variable $d_k$. So, the cost function is the total increase needed for all dominant cooling variables. Problem \eqref{new2} better captures the correlation between the servers.

We still need to present a heuristic for solving \eqref{new1} and \eqref{new2} because they are also NP-complete (problem \eqref{neww} is a special case for \eqref{new1} and \eqref{new2} in which there is only one cooling variable and $A$ is a constant matrix). The heuristic consists of three different phases, specified as follows.

Phase 1 is for rounding $\rho^*$ and generating $\hat{\rho}$. First note that $\rho^*$ is also the optimal solution for the LP relaxation of \eqref{new1} and \eqref{new2}, for which there is no need to increase the cooling variables. The idea is to gradually round the values in $\rho^*$ by considering the cost values for \eqref{new1} and \eqref{new2}.
 We know that the total workload for $\rho^*$ is $D$ and we need to find an integral solution with  total load $D$ for \eqref{new1} and \eqref{new2}. Let us denote such a workload distribution by $\hat{\rho}$. To go from $\rho^*$ to $\hat{\rho}$, some server workload modifications are required. We perform these modifications as follows. When a server that has positive load in $\rho^*$ has its load reduced to zero, its load is distributed among the remaining servers proportional to their current load. This way, the new distribution imitates the previous distribution.  If the resulting loads of some servers exceed 1, the extra load is again distributed among the remaining loaded servers. This is the base step in the following greedy heuristic.
  The heuristic selects the servers to be idle one by one. To select the next server to be idle, the heuristic successively reduces one of the $\rho_i>0$ to zero and by performing the same redistribution process a new distribution $\rho^{(i)}$ results. The cost for $\rho^{(i)}$ is then calculated. Finally, the server that is selected to be idle is the one that minimizes the cost of \eqref{new1} (or \eqref{new2}). This process is repeated until there are exactly $D$ servers with $\rho_i=1$. The procedure is shown in Algorithm 1. This gradual rounding can also be performed for problem \eqref{p1} directly. The issue is to calculate the cost function for \eqref{p1}, where a linear programming problem must be solved for each step. This increases the complexity of the heuristic as compared to problems \eqref{new1} and \eqref{new2}.

\begin{algorithm}
 \caption{Calculation of $\hat{\rho}$}\label{algorithm2}
  \begin{algorithmic}[1] 
  
    \State Solve the relaxed form of \eqref{p1} and call the solution $(v^*,\rho^*)$
   \State $S= \{i \in \rho^*| 0 <\rho_i \leq 1\}$
   \State $l=|S|-D$
   \State $\hat \rho=\rho^*$
   \While {$ l \neq 0$}
   \For {$i \in S$}
   \State $\rho^{(i)}=\hat\rho$
   \State $S'=S-\{i\}$
   \State $r= \rho^{(i)}_i $
   \State $\rho^{(i)}_i =0$
   \For {$j \in S'$}
   \State $\rho^{(i)}_j= \rho^{(i)}_j+ \frac{\rho^{(i)}_j}{\sum \limits_{k \in S'}\rho^{(i)}_k} r$
   \EndFor
\While {$\exists k \in S', \rho^{(i)}_k > 1$}
   \State $r=\rho^{(i)}_k -1 $
   \State $\rho^{(i)}_k =1$
\State $S'=S'-\{k\}$
  \For {$j \in S'$}
   \State $\rho^{(i)}_j= \rho^{(i)}_j+ \frac{\rho^{(i)}_j}{\sum \limits_{l \in S'}\rho^{(i)}_l} r$
   \EndFor
\EndWhile
\State $x_i=$ value of the cost function of \eqref{new1} (or \eqref{new2}) for $\rho^{(i)}$
   \EndFor
   \State Remove $i$ with the smallest $x_i$ from $S$ and $\hat \rho=\rho^{(i)}$
\State $l=l-1$
      \EndWhile
\State \textbf{return} $\hat{\rho}$
	 \end{algorithmic}
  \end{algorithm}

In Algorithm 1, the servers that are idle in $\rho^*$ remain idle. However, they may be working in the optimal integral solution. To allow these servers to work, in phase 2, after calculation of $\hat{\rho}$, there is another step that idles one of the working servers in $\hat{\rho}$ and then switches it with each of the servers that are idle. It finally chooses the server to work as the one that minimizes the cost of \eqref{new1} (or \eqref{new2}). For problem \eqref{new1} if more than one server minimizes the cost, then the server whose corresponding $\rho$  minimizes the summation of violations, $\sum_{l=1}^{n} \frac{(\sum_{i=1}^{n}B'_{l,i} \rho_i -(\sum_{j=1}^{m}A_{l,j} v^*_{j}+b-E_l) )^+}{w_l}$, is chosen. This final ranking is not needed for problem \eqref{new2}, because there is already a summation in the cost function.

Finally, in phase 3, the procedure explained so far is repeated for a specific number of iterations. At each iteration, a small perturbation of $A$, $B$ and $E$ is applied and the procedure is repeated. At the end, the solution that gives a lower cost for \eqref{p1} is chosen (by fixing $\hat{\rho}$, \eqref{p1} becomes a linear programming problem). The reason for these perturbations is the optimal fractional solution $\rho^*$ is more sensitive to small changes in $A$, $B$ and $E$ in comparison to the optimal integral solution. We assume these small changes do not affect the optimal integral solution but when starting from multiple initial $\rho^*$ values, we increase the chance of $\hat{\rho}$ values being close to the optimal integral solution by choosing the best solution ($\hat{\rho}$) among the perturbations.

\subsubsection{Genetic Algorithm}

Genetic algorithms are metaheuristics that try to search the solution space of a problem efficiently. They have been previously used in works closely related to our problem. A genetic algorithm is used in \cite{G3} for thermal-aware task scheduling. An enhanced genetic algorithm is proposed in \cite{G4} for reducing cooling power consumption.  

We also implement a genetic algorithm to see how such standard approaches work for this problem. 
The steps are similar to the algorithms presented in \cite{G1} and \cite{G2} for multiple knapsack and set covering problems, respectively.  
Each solution is represented by the 0-1 vector of workload distribution with size $n$. The costs are the value of the objective function for problem \eqref{p1}. There is intensification around the simple rounding solution when generating the initial population. The algorithm's phases are as follows.

\begin{itemize}
	\item \textbf{Population generation:} The initial population includes the simple rounding solution and the solutions where 10 percent of their ones (or zeros if the number of zeros is smaller, i.e., when $n-D<D$) are different from the simple rounding solutions. The perturbed utilizations are chosen uniformly at random.  

\item \textbf{Parent selection:} This is similar to \cite{G1}. It is a tournament with size two. Two sets of members with size two are chosen randomly from the population and the member with smaller cost from each set is chosen. 
\item \textbf{Child generation (cross over):} This is similar to \cite{G2}. In the places that both parents have 1 or 0, the child imitates them. If $f_1$ and $f_2$ are the cost values for parent1 and parent2, respectively, a proportion $\frac{f_2}{f_1+f_2}$ of the remaining ones are chosen from parent1 and a proportion $\frac{f_1}{f_1+f_2}$ are chosen from parent2.
\item \textbf{Mutation:}  According to \cite{G1}, one of the ones is randomly switched with one of the zeros.
\item \textbf{Replacement:} This is similar to \cite{G2}. The algorithm rates the solutions in terms of cost values. It then selects a solution uniformly from the second half of the rating and compares its cost with the cost of the child. If the child has a lower cost, it will replace the selected solution.

\end{itemize}

The parameters of the algorithm are the population size and the number of iterations for the last four phases in the algorithm. In Section \ref{eva}, we will specify these parameters.

\section{Evaluation}
\label{eva}

In this section, there are two main parts. In the first part, the performance of the proposed schemes will be evaluated for linear systems, in comparison to simple rounding and the proposed genetic algorithm. We evaluate performance while considering scalability and running time of the schemes. The performance is the ratio of calculated cost for the algorithms for problem \eqref{p1} to the optimal cost. The first set of systems is artificial, corresponding to the bad cases for simple rounding as explained in Section \ref{app}. These cases may be challenging for the algorithms, and so this set of experiments is designed to test the limits of the proposed heuristics. The second set of results corresponds to the model of the operational data center with 25 servers as described in Section \ref{sys}. We perform a regression on samples generated from the model of the operational data center to generate the linear system described by problem \eqref{p1}. To explore scalability, we generate two systems with sizes of 50 and 75 servers by scaling the matrices for the system with 25 servers. In the second part, for the model of the operational data center (the 25 server model), we evaluate using the approximate and exact solution of the linear system in the original system described by problem \eqref{P0}. The energy saving is evaluated by performing a comparison with the solution of the nonlinear system calculated by MATLAB. We also compare these results with the case of continuous server utilizations and a single upper bound for the server temperatures, as an example of how our proposed approach can be used to generate useful operational insights.

\subsection{Evaluation of the Proposed Schemes for Linear Systems}

The first two systems correspond to the two bad cases for simple rounding as explained in Section \ref{def}. For Case 1, there are 25 servers and three cooling variables. For the cooling matrix $A$, for each row the entries are 0, except for one that is chosen randomly, and its value is also uniformly distributed on the interval [0,1]. For the heat recirculation matrix $B$, each server is affected by five servers including itself. Thus there are five non-zero entries for each row of $B$, corresponding to the assumptions of Case 1. We set these five entries equal to 1. In this example, for the first row, the first five entries are equal to 1. For the second row we shift the ones to the right by one place. We continue in a similar manner for the remaining rows. To be precise:

\begin{equation}
	B_{i,j}=1 \Leftrightarrow j = (k+i-2 \quad mod \quad 25)+1, \quad k \in \{1,2,3,4,5\}
\end{equation}

So, there are five ones in each row and each column of $B$. We also set $T_{idle}=2$, $T_{busy}=1$, $V_{LB}=(10^{-3} \quad 10^{-3} \quad 10^{-3})$, $V_{UB}=(10^{8} \quad 10^{8} \quad 10^{8})$. With these settings, when $D \leq 5$ the optimal cooling values are $V_{LB}$, which is a condition for Lemma 1. $V_{LB}$ has small values to generate large approximation factors, as explained in the proof of Lemma 1. $V_{UB}$ has large values to guarantee the feasibility of the problem. In Case 2, we only change $A$ to correspond to the assumptions for Case 2. So, $A$ has equal rows whose entries are randomly chosen on the interval [0,1]. In this case, when $D \leq 5$, $V_{LB}$ is also the optimal cooling setting. Next, we generate Case 3 based on the previous cases to generate settings that may better reflect practice. Case 3 can be seen as a combination of Case 1 and Case 2 with smoother behaviour for $A$ and $B$. For $A$, the entries are randomly chosen to be 0, 1, 2 or 3 with the constraint that the summation of the values in each row is 3. For $B$, we assume each server is affected mostly by itself, which is a reasonable assumption in practice. We also assume that four other servers have higher effects on a specific server. So, we consider three uniform distributions. For each row, the entry in the main diagonal is chosen from the interval [2,5], there are four other entries randomly chosen from [1,2], and the other entries are selected from [0,0.5].

\begin{table*}
  \begin{center}
    \caption{Average Performance of the Algorithms Over 100 Runs for the Synthetic Linear Models}
    \label{t1}

    \begin{tabular}{|ccccccccccccccccc|} 
\hline
\multicolumn{17}{|c|}{Case 1}\\

\hline

&& \multicolumn{3}{c}{SR}&&\multicolumn{3}{c}{GA}&&\multicolumn{3}{c}{H1}&&\multicolumn{3}{c|}{H2} \\

\hline

$D$&&avg&wrc&pop&&avg&wrc&pop&&avg&wrc&pop&&avg&wrc&pop\\
\hline

4 && 
560 &  5688 &  0.20 &&  
36  &  375  &  0.90 &&  
51  &  792  &  0.87 &&  
1   &  1    &  1    \\

5 && 
901 &  5218 &  0.02 &&
334 &  546  &  0.13 &&    
408 &  1480  &  0.17 &&  
108 &  549  &  0.71   \\

6 && 
4.60 & 27.89&  0     && 
1.68 & 3.30 &  0.05  &&  
2.09 & 4.44 &  0.10  &&  
1.25 & 2.33 &  0.20     \\

7 && 
5.48 &  34.91 & 0    && 
1.95 &  3.73 & 0.05 &&  
2.07 &  5.09 & 0.05 &&  
1.41 &  3.05 & 0.17  \\

8 && 
12.84   &  248  & 0    &&  
1.97 &  4.10 & 0    && 
1.78 &  4.12 & 0.03 &&  
1.27 &  2.38 & 0.29 \\
 
 \hline

\multicolumn{17}{|c|}{Case 2}\\

\hline

4 && 
514 &  2015 &  0     &&  
64  &  553  &  0.84  &&  
1   &  1    &  1     &&  
1   &  1    &  1       \\

5 && 
509 &  1779 &  0     &&  
496 &  1779 &  0.02  &&  
472 &  1779 &  0.06  &&  
184 &  1121  &  0.62  \\

9 && 
2.16 &  3  &  0.02   &&  
1.55 &  2  &  0.45 &&  
1 .01   &  2  &  0.99    &&  
1    &  1  &  1    \\

10 && 
2.25 &  3  &  0     &&  
1.99 &  2  &  0.01  &&  
1.79 &  2  &  0.21  &&  
1.36 &  2  &  0.64  \\

11 && 
1.21 &  2   &  0.60    &&  
1 &  1 &  1 &&  
1 &  1 &  1 &&  
1 &  1 &  1 \\

 \hline

\multicolumn{17}{|c|}{Case 3}\\

\hline
 
1 && 
1.61 & 3.56 & 0.17  && 
1.17 & 2.04    & 0.47  &&  
1    & 1    & 1     &&  
1    & 1    & 1      \\
 
2 && 
1.79 &  3.66 & 0.04 &&   
1.26 &  2.37 & 0.24 &&  
1.31 &  2.75 & 0.31 &&   
1.12 &  2.04 & 0.47   \\
 
3 &&
1.60 &  3.03 & 0.02 &&  
1.24 &  2.06 & 0.13 &&  
1.35 &  2.79 & 0.12 &&  
1.11 &  1.64 & 0.36   \\
 
4 && 
1.34 &  2.47 & 0.11 && 
1.15 &  1.60 & 0.18 && 
1.21 &  1.70 & 0.14 && 
1.10 &  1.43 & 0.27   \\
 
5 && 
1.28 &  2.50 & 0.06 &&  
1.12 &  1.46 & 0.18 &&  
1.16 &  1.49 & 0.15 && 
1.09 &  1.45 & 0.25  \\
 
 \hline

   \hline

    \end{tabular}
  \end{center}
\end{table*}

Table \ref{t1} and Table \ref{t2} report the performance and running times of four algorithms for these different cases. The algorithms are simple rounding (SR), genetic algorithm (GA), proposed scheme based on solving problem \eqref{new1} (H1) and proposed scheme based on solving problem \eqref{new2} (H2). The results are the average over 100 runs of the algorithms for randomly generated systems. For GA, the population size and the number of iterations are set to $5 \times min(D, n-D)$ and $10 \times min(D, n-D)$, respectively. For H1 and H2, the number of perturbations, as explained in the last part of Section \ref{app}, is $min(5, D , n-D)$. In Table \ref{t1}, there are three metrics, the average ratio to the optimal cost calculated by MATLAB, avg, the worst case ratio to the optimal cost, wrc, and the proportion of times generating the optimal cost, pop. Due to lack of space we only report the results for five values of $D$ to highlight the performance differences between the algorithms. H2 has the best performance for the entire range of $D$. The performance for the middle range of $D$ ($5 \leq D \leq 15$) is more important because the search space for the solution is larger (and hence the decisions for how to reduce power consumption are more challenging). Table \ref{t2} reports the running times. Table \ref{t2} has an extra column, OPT, which reports the running time of solving the problem exactly in MATLAB (using the \textit{intlinprog} function). As explained in Section \ref{sys}, the platform was MATLAB R2021b running on a 64-bit system with an i7-1185G7 processor and 8-GB RAM.

According to Tables \ref{t1} and \ref{t2}, H2 has the best performance and a reasonable running time. For Case 1, as expected from Lemma 1, when $D \leq 5$ the approximation factor is large for SR. This is also true for Case 2, as expected from Lemma 2. In these two cases, the other algorithms work better and the best results are clearly for H2 throughout. For Case 3, H2 still has the best performance but with less of an advantage over the other algorithms. 
The results in Table \ref{t2} show the best running times are for SR and the worst are for GA. H1 and H2 have reasonable running times. The running times for H2 are more than for H1 and H1's running times scale more smoothly. The reason is H2 performs some calculations for determining the dominant cooling variables, along with the differences in the cost functions of \eqref{new1} and \eqref{new2}. Finally, from the results for Case 3 with 50 servers we can see that the MATLAB function cannot solve the problem efficiently. 

\begin{table}
  \begin{center}
    \caption{Average Running Time (in seconds) of the Algorithms Over 100 runs for the Synthetic Linear Models}
    \label{t2}

    \begin{tabular}{|ccccccc|} 
\hline
\multicolumn{7}{|c|}{Case 1}\\

\hline

$D$&& \multicolumn{1}{c}{SR}&\multicolumn{1}{c}{GA}&\multicolumn{1}{c}{H1}&\multicolumn{1}{c}{H2} & OPT\\

\hline

10 && 
0.018 &  1.221 &  0.115 &  
0.149  &  0.165     \\

11 && 
0.018 &  1.420 &  0.119 &
0.156 &  0.168     \\

12 && 
0.020 & 1.643 &  0.128    & 
0.159 & 0.160 \\

13 && 
0.020 &  1.676 & 0.128    &
0.163 &  0.136  \\

14 && 
0.019   &  1.538  & 0.122   &  
0.153 &  0.101 \\
 
 \hline

\multicolumn{7}{|c|}{Case 2}\\

\hline

11 && 
0.016 &  1.553 &  0.132    &  
0.212  &  2.555  \\

12 && 
0.020 &  1.691 &  0.127     &  
0.203 &  1.119  \\

13 && 
0.020 &  1.707  &  0.130   &  
0.203 &  0.552  \\

16 && 
0.021 &  1.278  &  0.130    &  
0.194 &  1.842   \\

17 && 
0.021 &  1.126   &  0.129    &  
0.195 &  0.464  \\

 \hline

\multicolumn{7}{|c|}{Case 3}\\

\hline
 
10 && 
0.017 & 1.160 & 0.108  & 
0.133 & 0.131         \\
 
11 && 
0.018 &  1.348 & 0.112 &   
0.139 &  0.146    \\
 
12 &&
0.018 &  1.513 & 0.114 &  
0.144 &  0.152   \\
 
13 && 
0.018 &  1.469 & 0.115 & 
0.147 &  0.156   \\
 
14 && 
0.018 &  1.337 & 0.111 &  
0.144 &  0.142 \\
 
 \hline

\multicolumn{7}{|c|}{Case 3 with 50 servers}\\

\hline

14 && 
0.022 & 1.933 & 0.195    & 
0.464 & 6.000  \\
 
15 && 
0.024 &  2.165 & 0.209   &  
0.508 &  8.183   \\
 
16 &&
0.024 &  2.319 & 0.213    &  
0.506 &  10.363 \\
 
17 && 
0.022 &  2.331 & 0.206    & 
0.486 &  11.744    \\
 
18 && 
0.023 &  2.534 & 0.213  &  
0.508 &  15.560  \\
   
   \hline

    \end{tabular}
  \end{center}
\end{table}

The other set of inputs comes from the linear regression of a model for the data center described in Section \ref{sys}. We also set $T_{idle}=35, T_{busy}=27, V_{LB}=(1300 \quad 10), V_{UB}=(2300 \quad 22)$. We performed linear regressions on the functions $M$ and $F$ using the \textit{regress} function in MATLAB. The data points are chosen uniformly at random from the defined ranges for the cooling variables and server untilizations (utilizations are used for regression of $M$ and they are continuous in this case). After performing the modifications explained in Section \ref{def}, the standard form of problem \eqref{p1} is generated. For evaluation of scalability, we change the cooling and heat recirculation matrices as follows. We consider three systems with 25, 50 and 75 servers. For 25 servers, instead of one cooling variable corresponding to total air flow of the fans, we assume that each fan has a separate air flow that affects closer servers more. For example, the leftmost servers are affected 90 percent by the left fan and 10 percent by the right fan. By considering these changes, we obtain a new cooling matrix with size $25 \times 3$. For the case of 50 and 75 servers both the cooling and heat recirculation matrices change. The details of how these changes are made are provided in the next paragraph.

\begin{table*}
\label{t3}
  \begin{center}
    \caption{Average Performance of the Algorithms Over 100 Runs for the Linear Data Center Models}
    \label{t3}

    \begin{tabular}{|ccccccccccccccccc|} 
\hline
\multicolumn{17}{|c|}{Data center with 25 servers}\\

\hline

&& \multicolumn{3}{c}{SR}&&\multicolumn{3}{c}{GA}&&\multicolumn{3}{c}{H1}&&\multicolumn{3}{c|}{H2} \\

\hline

$D$&&avg&wrc&pop&&avg&wrc&pop&&avg&wrc&pop&&avg&wrc&pop\\
\hline

10 && 
1.03 &  1.08 &  0.10 &&  
1  &  1.02  &  0.32 &&  
1  &  1.03  &  0.45 &&  
1   &  1.02    &  0.38    \\

13 && 
1.04 &  1.07 &  0.02 &&
1.02 &  1.05  &  0.05 &&    
1 &  1.05  &  0.54 &&  
1 &  1.04  &  0.55    \\

14 && 
1.03 & 1.06 &  0    && 
1.02 & 1.05 &  0.03  &&  
1.01 & 1.05 &  0.30  &&  
1.01 & 1.05 &  0.31     \\

15 && 
1.03 &  1.07 & 0.01    && 
1.01 &  1.05 & 0.13 &&  
1.01 &  1.05 & 0.16 &&  
1.01 &  1.05 & 0.14 \\

 \hline

\multicolumn{17}{|c|}{Data center with 50 servers}\\

\hline

11 && 
1 &  1.01 &  0.13     &&  
1  &  1.01  &  0.15  &&  
1   &  1.01    &  0.16 &&  
1   &  1.01    &  0.17  \\





 \hline

\multicolumn{17}{|c|}{Data center with 75 servers}\\

\hline
 
16 && 
1.02 & 1.06 & 0.03  && 
1.01 & 1.01 & 0.04 &&  
1.01 & 1.01  & 0.10 && 1.01 & 1.01  & 0.11      \\

 \hline

   \hline

    \end{tabular}
  \end{center}
\end{table*}

For scaling from 25 servers to 50 servers, we consider two sets of five racks (each consisting of 25 servers) adjacent to each other. Instead of two fans, there are three fans, one between the two sets of racks, with two more at either end. We assume the airflow of the middle fan is spread equally between the two sets of racks and each set is affected by the fans at each of its ends. This way we can construct the cooling matrix (we did not change the coefficients for the chiller). For the heat recirculation matrix, the effects of servers in the two sets of racks on each other are considered as follows. The leftmost rack in the left set of racks is affected the least by the right set of racks. We assume the effects are 15 percent of the effects of the servers in the same set of racks. For the next rack (the second rack from the left) the effects are similar but larger due to the shorter distance. So, we assume the effects are twice the effects calculated for the leftmost rack. For the next racks, the effects are three, four and five times the effects calculated for the leftmost rack, respectively. Finally, for the case of 75 servers, we assume the racks that are not adjacent are isolated from each other in terms of heat recirculation.

Table \ref{t3} reports the results. The problem is solved 100 times. Each time the matrices $A$, $B$ and $E$ are perturbed by multiplying each of their entries with a random value drawn from a uniform distribution on the interval [0.98, 1.02]. In these practical settings, the results for the different algorithms are much closer compared to the synthetic cases. However, H1 and H2 still outperform SR and GA. While we ran a more extensive set of experiments, here we only report a few values of $D$ that highlight the differences. Similarly to the results in Table \ref{t2}, we also checked the running times in this case and scalibility was also confirmed in terms of running times.

\subsection{Evaluation of Energy Savings for the Original System}

We now evaluate the effectiveness of our proposed approach versus attempting to compute the optimal solution directly. We solve problem \eqref{P0} with the actual nonlinear functions $M$ and $F$ in MATLAB (using the \textit{surrogateopt} function) and then compare the results with the results obtained by using the solution of the linearized system. The case considered corresponds to the operational data center of 25 servers and two cooling variables as described in Section \ref{sys}.  In Table \ref{t4}, there are three extra columns, OPTL, OPT27 and OPT. OPTL and OPT correspond to solving the linear problem and solving problem \eqref{P0}, respectively, both in MATLAB. We explain OPT27 in the next paragraph. All the solutions are evaluated for the original system. We also checked the feasibility of the solution for the original system. The results reported in Table \ref{t4} are the ratio to the best cost value found by the algorithms for each value of $D$. As reported in the last two columns and described in Section \ref{sys}, the MATLAB function with default settings cannot return the optimal solution in most of the cases. It is also very time consuming due to evaluation of the function $M$ at each iteration as explained in Section \ref{sys}. The results show that the linear regression and the corresponding approximation schemes, H1 and H2 in particular, have close to the best performance in the middle range of $D=2$ to $D=20$.  Overall, for the middle range of $D$ that is more problematic to solve and more common in practice, the proposed schemes show the best performance (up to 45 percent reduction in the power consumption compared to the solution calculated with MATLAB).

Finally, to investigate the effect of considering two different upper bounds for the temperatures of idle and busy servers, we report the results for the case of continuous utilizations but with one upper bound for the server temperatures equal to $T_{busy}$ (we have to choose the smallest upper bound, $T_{busy}$, because it should work for all utilizations). In Table \ref{t4}, column OPT27 corresponds to this case by setting $T_{busy}=27$ and solving the problem in MATLAB (using the \textit{fmincon} function that is also slow due to the need to repeatedly evaluate the function $M$). As shown, the case of continuous utilizations does not yield acceptable performance for values of $D$ up to $D=20$. However, it has the best performance for the largest values of $D$. The results also show that we can take advantage of the two different red-line temperatures to reduce the total power consumption significantly compared to the solution calculated with MATLAB for the case of a single red-line temperature (by up to 30 percent for the middle range of $D$).

\begin{table}
  \begin{center}
    \caption{Performance of Different Algorithms with Results Applied to the Original 25-server System}
    \label{t4}

    \begin{tabular}{|ccccccccc|} 
\hline
\multicolumn{9}{|c|}{Original 25-server system}\\

\hline

$D$&& \multicolumn{1}{c}{SR}&\multicolumn{1}{c}{GA}&\multicolumn{1}{c}{H1}&\multicolumn{1}{c}{H2} & OPTL & OPT27 & OPT\\

 \hline

1 && 
1.27 & 1.17 & 1.17    & 
1.17 & 1.17 & 1.47 & 1 \\

2 && 
1 & 1 & 1   & 
1 & 1 & 1.27  & 1 \\

3 && 
1 & 1 & 1   & 
1 & 1 & 1.28 &  1\\

4 && 
1.08 & 1 & 1.01   & 
1 & 1 &1.30 &  1\\

5 && 
1 & 1 & 1    & 
1 & 1 &1.30  & 1.10 \\

6 && 
1.01 & 1.01 & 1.01   & 
1.01 & 1.01 & 1.32 & 1 \\

7 && 
1 & 1 & 1    & 1
 & 1 & 1.31 & 1  \\

8 && 
1.21 & 1 & 1.01    & 
1.01 & 1 & 1.30 & 1 \\

9 && 
1.10 & 1.10 & 1   & 
1 & 1 & 1.29 &  1.05 \\

10 && 
1.09 & 1.09 & 1   & 
1.09 & 1 & 1.28 & 1.06  \\

11 && 
1.03 & 1.02 & 1    & 
1.02 & 1 & 1.17 & 1.20 \\

12 && 
1.01 & 1 & 1    & 
1 & 1 & 1.15 & 1.16 \\

13 && 
1.02 & 1.02 & 1   & 
1 & 1 & 1.16 & 1.06 \\

14 && 
1.02 & 1 & 1    & 
1 & 1 & 1.16 & 1.43 \\

15 && 
1 & 1 & 1    & 
1 & 1 & 1.17 & 1.46 \\

16 && 
1.01 & 1.01 & 1   & 
1 & 1 & 1.16 & 1.45 \\

17 && 
1.02 & 1 & 1   & 
1 & 1 & 1.14 & 1.43 \\

18 && 
1.01 & 1.01 & 1   & 
1 & 1 & 1.13 & 1.40 \\

19 && 
1 & 1 & 1    & 
1 & 1 & 1.11 & 1.37 \\

20 && 
1 & 1 & 1    & 
1 & 1 & 1.09 & 1.36 \\

21 && 
1.51 & 1.45 & 1.40   & 
1.40 & 1.39 & 1 & 1.18 \\

22 && 
1.46 & 1.40 & 1.36   & 
1.36 & 1.36 & 1  & 1.13 \\

23 && 
1.40 & 1.35 & 1.33   & 
1.33 & 1.33 & 1 &  1.08 \\

24 && 
1.35 & 1.31 & 1.31    & 
1.31 & 1.31 & 1 & 1.07 \\

   \hline

    \end{tabular}
  \end{center}
\end{table}

\section{Conclusion}
\label{con}
In this paper, we introduced a problem considering several cooling parameters, heat recirculation effects and two red-line temperatures for servers with a goal of minimizing the power consumption for a data center. We proposed a framework including some assumptions, linearization of the problem and a heuristic for intelligent rounding of the fractional solution for the linearized model. We provided an analysis of the optimization problem for the linearized problem, by introducing simple rounding and genetic algorithms as the baseline algorithms and extracting two bad cases for simple rounding that are also used for evaluation of the proposed schemes. The proposed framework showed acceptable performance and running times for the cases of synthetic and real world systems. The solution of the linear system is also well suited for the real system. In some cases, the power consumption for the real system is reduced by 45 percent when compared with the solutions returned by MATLAB, which requires a much longer running time. The results also confirm that considering two red-line temperatures is beneficial to reduce the cooling power consumption by up to 30 percent (which also has much more running time as compared to the proposed heuristics). This was demonstrated by comparison with the case of one red-line temperature and continuous utilizations solved with MATLAB. As for future work, the applicability of linearization for the case of server consolidation can be investigated. This work may also be a good starting point for considering deeper analysis of the underlying optimization problems. For example, the case of heterogeneous data centers is a more complicated problem that can be investigated. Additional costs such as migration cost (moving workload between servers) can also be considered, when one considers a problem where the workload varies over time.

\begin{IEEEbiography}[{\includegraphics[width=1in,height=1.25in,clip,keepaspectratio]{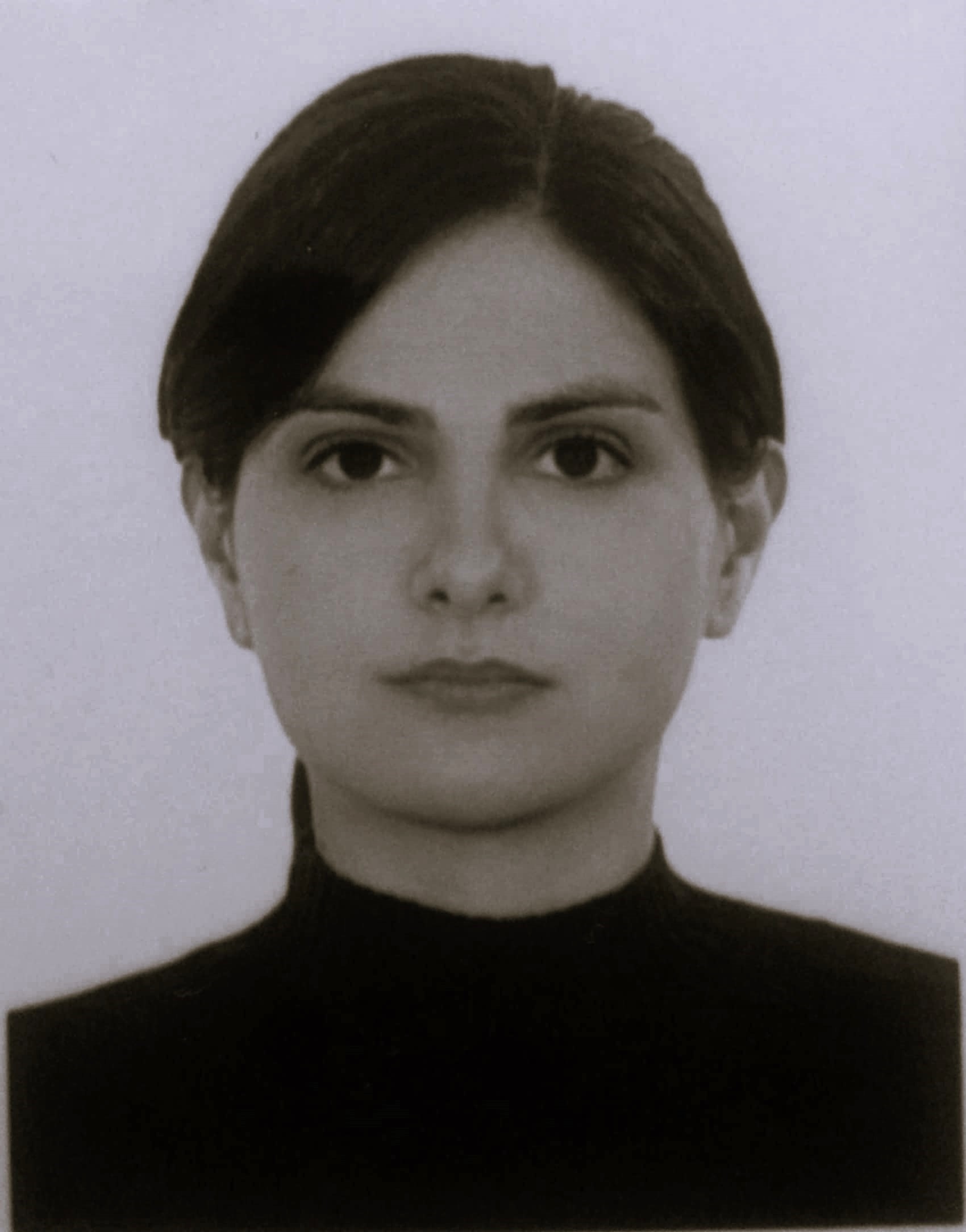}}]{Somayye Rostami}
received her B.A.Sc. and M.A.Sc.
degrees in electrical engineering from the University of Tehran (2012) and the Iran University of Science \& Technology (2015), Tehran, Iran, respectively. She is a PhD student in computer science at McMaster University, Hamilton, ON, Canada. Her research interests are optimization, algorithms and stochastic analysis with applications in resource allocation, energy consumption, economics and networks.


\end{IEEEbiography}

\begin{IEEEbiography}[{\includegraphics[width=1in,height=1.25in,clip,keepaspectratio]{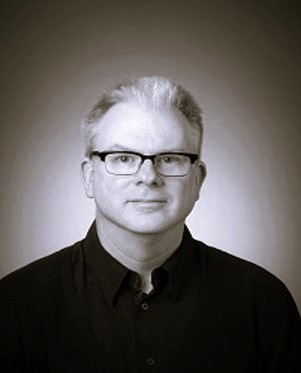}}]{Douglas G. Down}
 (SM’05) received the B.A.Sc. and M.A.Sc. degrees in electrical engineering from the University of Toronto, Toronto, ON, Canada, in 1986 and 1990, respectively, and the Ph.D. degree in electrical engineering from the University of Illinois at Urbana-Champaign, Urbana-Champaign, IL, USA, in 1994. He is a Professor in the Department of Computing and Software, McMaster University, Hamilton, ON, Canada. His interests lie in scheduling and performance evaluation for distributed systems, workload and thermal management for data centers, and scheduling of systems with predictions.

\end{IEEEbiography}

\begin{IEEEbiography}[{\includegraphics[width=1in,height=1.25in,clip,keepaspectratio]{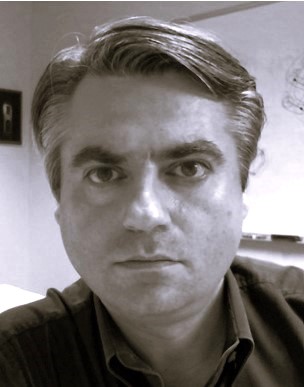}}]{George Karakostas}
 is an associate professor in the Department of
Computing \& Software of McMaster University, in Hamilton ON,
Canada. He holds an Eng. Diploma in Computer Engineering \& Informatics
from the Univercity of Patras, Greece, and MASc and PhD degrees in
Computer Science from Princeton University. His main research
interests are in the field of Theoretical Computer Science, and, more
specifically, in the design and analysis of approximation algorithms,
algorithmic game theory, and their applications in practical fields
such as energy consumption and networks.


\end{IEEEbiography}







\end{document}